\documentstyle[12pt]{article}
\newcommand{\fnd}[2]{\frac{\textstyle #1}{\textstyle #2}}
\newcommand{\xrm}[1]{{\textstyle \mbox{\rm #1}}}
\newcommand{\bm}[1]{\mbox{\boldmath $#1$}}
\newcommand{\abs}[1]{\left| #1\right|}

\begin{document} \baselineskip .7cm
\title{\bf On the dominance \\
of non-exotic meson-meson scattering \\
by \bm{s}-channel \bm{q\bar{q}} confinement states \\
and \\ the classification of the scalar mesons}
\author{{\large Eef van Beveren{\normalsize $^{\; a}$} and
George Rupp{\normalsize $^{\; b}$}}\\ [.5cm]
{\normalsize\it $^{a\;}$Centro de F\'{\i}sica Te\'{o}rica, Departamento de
F\'{\i}sica,}\\ {\normalsize\it Universidade, P3004-516 Coimbra, Portugal,}
{\small (eef@teor.fis.uc.pt)}\\ [.3cm]
{\normalsize\it $^{b\;}$Centro de F\'{\i}sica das Interac\c{c}\~{o}es
Fundamentais, Instituto Superior T\'{e}cnico,}\\ {\normalsize\it
Edif\'{\i}cio Ci\^{e}ncia,
P1049-001 Lisboa Codex, Portugal,} {\small (george@ajax.ist.utl.pt)}\\ [.3cm]
{\small PACS number(s): 11.80.Et, 12.40.Yx, 13.75.Lb, 14.40.-n}\\ [.3cm]
{\small hep-ex/0110156}\\ [.5cm]
{\normalsize Prepared for the Proceedings of the}\\
{\normalsize IX International Conference on Hadron Spectroscopy, HADRON 2001}\\
{\normalsize Protvino (Russian Federation) Aug 25 -- Sept 1, 2001}
}
\maketitle

\begin{abstract}
Non-exotic scalar-meson resonances in $S$-wave meson-meson scattering are
studied in the light of a unitarised Schr\"{o}dinger model.
The resulting poles in the scattering matrices, by analytical
continuation into the complex-energy plane, are grouped into nonets of
isoscalar, isodoublet, and isotriplet resonances.
All singularities can be related to {\it quark-antiquark confinement states},
the light-quark nonet of which has ground states at 1.3 to 1.4~GeV and level
spacings of some 300--400~MeV, except for a nonet of light scalar mesons below
1~GeV. All non-exotic $S$-wave resonances reported by experiment fit into this
scheme.
\end{abstract}

\section*{Introduction}

Lattice QCD in principle offers the most direct way to link to experiment
what we believe to be the fundamental theory of strong interactions
\cite{lQCD}.
However, in view of the constant evolution of results from sophisticated
lattice QCD solutions \cite{qlQCD}, it becomes ever more puzzling why
mesonic resonances can be described as simple quark-antiquark systems
in effective theories \cite{Ricken00}.
But apparently, such a scenario works!

Seemingly, the perturbative vacuum states of QCD at low energies are not
quarks and gluons, but rather confined constituent quarks and residual
interactions, a picture that has been successful for several decades by now.
What do lattice calculations teach us about constituent quarks?
One might think of colour-triplet configurations of quarks, antiquarks,
and glue, maybe with admixtures of higher colour multiplets, which mutually
feel colour interactions. Can lattice QCD identify such substructures?
Moreover, what happens exactly when those substructures suddenly turn
colourless and cease to be confined?
Does one observe colourless substructures that drift apart on the lattice?

In unitarised meson models one assumes an effective mass for the constituent
quark, a confinement force for the remaining colour interactions, and
a mechanism for decay \cite{unitarisation,ccbb80}.
However, not knowing how the separation into constituent quarks, confinement,
and decay can be derived from QCD, each model is the result of educated
guesses, rather than of rigorous derivations starting from QCD.
This frustrating state of the affairs has in the past three decades led to
a proliferation of effective models and theories.
However, no model exists so far which completely describes the resonances
of meson-meson scattering to a satisfactory degree of accuracy.
Nevertheless, some educated guesses are less successful than others.
For example, $q\bar{q}$ models for resonances that do not take meson
loops into consideration \cite{Ricken00} can never be in agreement with
experiment, since the resulting spectrum consists of zero-width bound states,
whereas large widths are measured.

\section*{Non-exotic meson-meson scattering}

The assumption that non-exotic meson-meson scattering is dominated by
the $s$-channel states of the confinement mechanism has been worked
out in a series of papers \cite{ccbb80,BRMDRR86,BRRD83,NUMM}.
Here, we will confine our attention to a toy model that we studied
in Ref.~\cite{BRBW01}.
There we obtain for the elastic low-energy partial-wave meson-meson
scattering matrix,

\begin{equation}
S_{\ell}(p)\; =\;\exp\left( 2i\delta_{\ell}(p)\right)
\;\;\; ,
\label{Smat}
\end{equation}

\noindent
the relation

\begin{equation}
\xrm{cotg}\left(\delta_{\ell}(p)\right)\; =\;
\fnd{n_{\ell}(pa)}{j_{\ell}(pa)}\; -\;
\left[ 2\lambda^{2}\;\mu\; pa\; j^{2}_{\ell}(pa)
\sum_{n=0}^{\infty}
\fnd{\abs{{\cal F}_{n}(a)}^{2}}{E-E_{n}}\right]^{-1}
\;\;\; ,
\label{deltaBW}
\end{equation}

\noindent
where $p$ represents the relative momentum in the CM frame of the two
mesons, $\ell$ their relative angular momentum, and $\mu$ their reduced mass;
$E_{n}$ ($n=0$, 1, 2, $\dots$) represent the energy eigenvalues of the
constituent $q\bar{q}$ system to which the meson pair couples, and
${\cal F}_{n}$ the corresponding $q\bar{q}$ eigenfunctions;
$n_{\ell}$ and $j_{\ell}$ stand for the spherical Bessel
and Neuman functions, respectively.

The intensity of the coupling between the meson-meson system and the
$q\bar{q}$ system is described by the parameter $\lambda$,
whereas $a$ stands for the average distance at which the transitions from
one system to the other take place \cite{Bev84}.

For $\lambda =0$, we find $S_{\ell}(p)=1$, which describes a system of two
non-interacting mesons. For small values of $\lambda$, the scattering
matrix (\ref{Smat}) has poles in the lower half of the complex-energy
plane, which can approximately be given by

\begin{equation}
E_\xrm{pole}\;\approx\; E_{n}\; -\;\abs{{\cal F}_{n}(a)}^{2}\;\left[
\sum_{n'\neq n}
\fnd{\abs{{\cal F}_{n}(a)}^{2}}{E_{n}-E_{n'}}\; -\;
\fnd{i}{2\lambda^{2}\;\mu\; pa\; j_{\ell}(pa)\; h^{(1)}_{\ell}(pa)}
\right]^{-1}
\;\;\; ,
\label{singapprox}
\end{equation}

\noindent
indicating that to each value of the radial quantum number $n$
corresponds one singularity, {\it i.e.}, one meson-meson scattering resonance.
For higher values of $\lambda$, one can determine the locations of the
poles in the scattering matrix by numerical methods.

The real parts of the singularities roughly correspond to the central
resonance positions, $E_{r}$, whereas the moduli of the imaginary parts
approximately equal half the resonance widths, $\Gamma_{r}$. In short, 

\begin{equation}
E_\xrm{pole}\;\approx\; E_{r}\; -\; i\;\fnd{\Gamma_{r}}{2} \; .
\label{cenwid}
\end{equation}

\noindent
Singularities may be located on the real energy axis below threshold,
representing stable (with respect to strong decay) mesons
({\it e.g.} $K$, $J/\Psi$, $\Upsilon$ \cite{BRRD83}).

For practical purposes, one might truncate the sum in formula
(\ref{deltaBW}) and substitute the truncated part by a constant \cite{BRBW01}.

For $\lambda\rightarrow\infty$ one finds

\begin{equation}
\xrm{cotg}\left(\delta_{\ell}(p)\right)\; =\;
\fnd{n_{\ell}(pa)}{j_{\ell}(pa)}
\;\;\; ,
\label{deltaBWinfty}
\end{equation}

\noindent
which represents scattering from a hard sphere of radius $a$. In this
case, the interior of the $q\bar{q}$ state becomes unobservable and no
resonance spectrum can be deduced from meson-meson scattering.

\section*{Comparison with experiment}

In Ref.~\cite{BRBW01} we compare the predictions of formula (\ref{Smat}) for
$K\pi$ $S$- and $P$-wave scattering in $I=1/2$ with the experimental cross
sections. For $P$ waves, in the region of the
$K^{\ast}$(892) resonance, we find that the lowest-lying state of the
confinement spectrum is at some 945~MeV, whereas the corresponding
pole comes out at $(887-27i)$~MeV.

In a more refined model \cite{BRRD83}, which also takes inelasticity
into account by considering the coupling to all channels with allowed
initial and final states of pseudoscalar and vector mesons, and furthermore
employs a more sophisticated mechanism for the coupling of $q\bar{q}$
confinement states to meson-meson scattering channels,
it is found that the ground state of the confinement spectrum in this
case comes out at 1.19~GeV, some 300~MeV above the position of the
$K^{\ast}$(892) pole.
It shows that bare states can be several hundreds of MeVs away
from the actual central resonance positions, and, moreover, that such
conclusions are model dependent. The toy model of Ref.~\cite{BRBW01} yields
a shift of only some 60~MeV. In this perspective, the question as to
where a bound-state model should find its {\it bare} \/states is hard to be
answered.

The latter question becomes even more difficult in the case of
$S$-wave scattering. There we find, in the toy model of Ref.~\cite{BRBW01},
that the ground state of the confinement spectrum is at 1.31~GeV and
the corresponding pole at $(1.46-0.12i)$~GeV. However, further
inspection of the singularity structure of the scattering matrix reveals
another pole far below this energy region, namely at $(714-228i)$~MeV.
The latter singularity has no direct relation to any of the bare states.
Hence, the $I=1/2$, $J^{P}=0^{+}$ ground state of bound-state models
should be at some 1.3~GeV and not below 1~GeV.
This result is confirmed by the full model \cite{BRMDRR86}
( pole at $(727-263i)$~MeV), in which also the
poles belonging to the two isoscalars $f_{0}(980)$ and the rather
controversial $f_{0}(470-208i)$ ($\sigma$ meson), as well as to the isovector
$a_{0}(980)$, have no direct relation with the ground states of the
corresponding confinement spectrum at about 1.2~GeV.

\section*{A model study of poles}

The relation between the $q\bar{q}$ bare states and the poles in
the corresponding scattering matrix can be found in the coupled-channel
model (\ref{Smat}) by considering the process of stepwise reducing the
coupling constant $\lambda$. Such a study has been performed in detail in
the toy model of Ref.~\cite{BRBW01}, with the following result.
All poles move towards a corresponding $q\bar{q}$ bare state on
the real axis, as predicted by formula (\ref{singapprox}), except for
the $S$-wave singularity below 1~GeV. The negative imaginary part
of this pole grows inversely proportionally to $\lambda^{2}$,
implying that the corresponding ``resonance'' disappears into the
background of $K\pi$ scattering. Unfortunately, such processes cannot be
tested in experiment, since Nature corresponds to a fixed value for
$\lambda$.

\section*{Constituent-quark-pair creation}

For $P$- and higher-wave meson-meson scattering, we do not find singularities
other than those which can be related to a $q\bar{q}$ state of the
confinement spectrum. The extra poles below 1~GeV, described
through {\it pole doubling} \/in Ref.~\cite{BR99a}, exclusively appear in
$S$-wave meson-meson scattering. Hence, we must conclude that the latter
``resonances'' are a consequence of the mechanism of constituent-quark-pair
annihilation and creation, which couples meson-meson initial and final states
to the $q\bar{q}$ confinement states.

For $P$ and higher waves, the centrifugal barrier prevents the formation of
such resonances in meson-meson scattering. But in the absence of a centrifugal
barrier for $S$ waves, resonances are formed that in the cases of the
$f_{0}(980)$ and the $a_{0}(980)$ are narrow enough to be clearly observable,
but which for the $f_{0}(470-208i)$ and $K^{\ast}_{0}(727-263i)$ are
too broad to be firmly established.

One should note here that, when a pole with a large imaginary part
lies close to threshold --- close meaning that the distance from threshold to
the real part of the singularity is smaller than or of the same order
as the imaginary part --- then the corresponding cross section has a shape
which is very different from a standard Breit-Wigner.
In the Argand plot, one finds a resonance motion that rapidly slows down for
higher energies. If then, moreover, new thresholds get open and other rapid
Breit-Wigner resonances show up, its appearance can hardly be recognised as
that of a resonance, within the experimental accuracy.

Nevertheless, whether or not one associates a resonance with the controversial
$J^{P}=0^{+}$ singularities is of {\it no importance}.
What is crucial in the above observations is the fact that it settles the
classification of the $f_{0}(980)$ and $a_{0}(980)$ resonances in a
nonet scheme for mesons rather more naturally than other proposals.

\section*{Resonance shapes}

The model result that some of the light scalar resonances are broad,
while others are narrow, has its origin in the effects of inelasticity.
In Ref.~\cite{BRMDRR86}, Table 1 of the Appendix, a list of
inelasticity channels for the three scalar-meson isomultiplets is presented, as
well as the intensities of the relative couplings.

We learn from this table that the isotriplet couples twice as strongly to
$\eta_{n}\pi$ as to $K\bar{K}$, other thresholds lying at higher or much
higher energies, which makes their effect hardly relevant here.
However, only a small part of the $\eta_{n}\pi$ channel decays into
$\eta\pi$, the rest into $\eta'\pi$. This implies that, of the lowest
channels, the $K\bar{K}$ and also the $\eta'\pi$ channel are far stronger
than the $\eta\pi$ channel (see also Ref. \cite{Oset}).
Elastic $S$-wave $\eta\pi$ scattering in the absence of inelasticity can be
described by the toy model of formula (\ref{deltaBW}).
In Ref.~\cite{BRBW01}, formula (\ref{deltaBW}) has been applied to 
elastic $S$-wave $K\pi$ scattering.
Now, when we substitute there the $K$ mass by the $\eta$ mass, then we
obtain a toy model for elastic $S$-wave $\eta\pi$ scattering.
With this substitution, we find for the model of formula (\ref{deltaBW})
indeed a pole close to the $\eta\pi$ threshold, and with a relatively
large imaginary part ($763-199i$~MeV).
Moreover, the related toy-model prediction for the $\eta\pi$
elastic $S$-wave scattering cross section does not show a clear resonance,
exactly as in the case of the $K^{\ast}_{0}(727-263i)$ pole in $K\pi$
elastic scattering.

Inelasticity, which has been taken into account in Ref.~\cite{BRMDRR86}, has
two consequences here:
first, the pole moves close to the $K\bar{K}$ threshold, with a smaller
imaginary part ($968-28i$~MeV), and, second, since the $\eta\pi$
threshold is far enough below that pole, its resonance shape turns
more Breit-Wigner-like (see {\it e.g.} \/Fig.~2 of Ref.~\cite{BRMDRR86}).
Nevertheless, upon reducing the coupling constant, the modulus of the
imaginary part of this pole increases in a similar way as does the lower
pole, the $K^{\ast}_{0}(727-263i)$, in $K\pi$ elastic $S$-wave scattering.
Consequently, the two poles have a similar origin, not directly related
to the bare spectrum. It is only through the strong interference of
the $K\bar{K}$ and the $\eta'\pi$ channels that a reasonable
Breit-Wigner-like shape appears for the light isotriplet resonance
$a_{0}(980)$.

We also observe from the above-referred table of Ref.~\cite{BRMDRR86}
that, of the isoscalar complex ($n\bar{n}$ coupled to $s\bar{s}$),
the $n\bar{n}$ couples strongly to $\pi\pi$, whereas the $s\bar{s}$
couples strongly to $K\bar{K}$, with, again, other thresholds lying higher
or much higher, thus making their influence of little importance here.
Furthermore, $n\bar{n}$ and $s\bar{s}$ are coupled to one another through
the $K\bar{K}$ channel, which implies that the $s\bar{s}$ component of
the isoscalar complex also couples to $\pi\pi$, but quite weakly.
Hence, for the $s\bar{s}$ resonance $f_{0}(980)$ we can now repeat the
arguments we gave for the isotriplet resonance shape, with $\eta\pi$
replaced by $\pi\pi$. Moreover, since the $\pi\pi$ threshold in the isoscalar
case lies much lower than the $\eta\pi$ threshold in the isotriplet case, we
find a more convincing Breit-Wigner-like shape for the $f_{0}(980)$ in
$\pi\pi$ scattering \cite{DM2} than for the $a_{0}(980)$ in $\eta\pi$
scattering.

However, the $n\bar{n}$ component of the isoscalar complex, which yields a
pole close to the $\pi\pi$ threshold, has no further strong-inelasticity
channel to allow for a Breit-Wigner-like shape for the corresponding
resonance $f_{0}(470-208i)$.
The same happens to the isodoublet, which, according to the table of
coupling constants mentioned before, couples strongly to the $K\pi$ channel.
No further lower-lying inelasticity channel exists in this case.
Consequently, also the $K^{\ast}_{0}(727-263i)$ has no
Breit-Wigner-like shape.

None of the poles of this nonet of scalar {\it resonances} \/has a direct
relation to the bare spectrum.
By stepwise reducing the model coupling constant, all nine poles stepwise
disappear into the complex plane with increasing negative imaginary part,
whereas the corresponding structures in the meson-meson scattering cross
sections stepwise disappear into the background.

\section*{The \bm{K}-matrix}

As one can easily observe from formula (\ref{deltaBW}), we have no
poles in the $K$-matrix at the energy eigenvalues $E_{n}$
of the confinement spectrum, since the hard-sphere-scattering
part in the expression for the cotangent of the phase shift does
not vanish at energies $E_{n}$.
This contradicts the observation of Sarantsev and collaborators
(Ref.~\cite{ANS01} and references therein)
that bare states are the singularities of the $K$-matrix, so this issue
deserves further study.

In the limit of an infinitely strong coupling between the confinement
and scattering sectors, formula (\ref{deltaBW}) predicts
that no bare spectrum can be observed in meson-meson scattering
other than the hard-sphere spectrum. This is reasonable, since in that
limit the mesons become impenetrable and thus do not allow the
observation of the interior dynamics.
However, when the hard-sphere-scattering part of formula (\ref{deltaBW})
is removed, then one just obtains stronger resonances close to the
poles of the $K$-matrix when the coupling constant $\lambda$ is
increased.

For small coupling, both models give similar results, except that the real
shifts for formula (\ref{deltaBW}) can be much larger when the 
hard-sphere-scattering part is present.
This is probably the reason why the bare states of Ref.~\cite{ANS01} are always
close to the central resonance energies.

We may therefore conclude that the behaviour of both models for moderate
coupling is very similar, except for the interpretation of the bare states.
In Ref.~\cite{extended}, we studied other consequences of the fact
that mesons are not point particles, but finite distributions of
constituent quarks.

Nevertheless, the fits of Ref.~\cite{AS97} to the data are too good
for the corresponding model to be totally wrong.
It might be that, with a small modification, the latter model would also yield
the extra $J^{P}=0^{+}$ nonet of singularities and no related
bare states, without destroying the excellent fits to the data.

\section*{\bm{J^{P}=0^{+}} resonances nonets}

In Table~\ref{nonets}, we classify \cite{BRMDRR86}
the experimentally observed non-exotic scalar mesons into nonets.

\begin{table}[ht]
\begin{center}

\begin{tabular}{|c||c|c|c|}
\hline & & & \\ [-0.3cm]
radial & & & \\
excitation & isotriplets & isodoublets & isoscalars \\ [.2cm]
\hline\hline & & & \\ [-0.3cm]
pole doubling & $a_{0}(980)$ & $K^{\ast}_{0}(727-263i)$ &
$f_{0}(470-208i)$ and $f_{0}(980)$ \\
[.2cm] \hline & & & \\ [-0.3cm]
ground state & $a_{0}(1470)$ & $K^{\ast}_{0}(1430)$ &
$f_{0}(1370)$ and $f_{0}(1500)$\\
[.2cm] \hline & & & \\ [-0.3cm]
first & & $K^{\ast}_{0}(1950)$ &
$f_{0}(1710)$ and $f_{0}(?)$\\
[.2cm] \hline & & & \\ [-0.3cm]
second & & & $f_{0}(2020)$ and $f_{0}(2200)$\\
[.2cm] \hline\end{tabular}
\end{center}
\caption[]{The nonet classification \cite{BRMDRR86} of
the $S$-matrix poles for $J^{P}=0^{+}$ meson-meson scattering.}
\label{nonets}
\end{table}

The $f_{0}(470-208i)$ comes in the {\it Tables of Particle Properties},
Ref.~\cite{PP}, under $f_{0}(400\mbox{--}1200)$,
but is not well established as a resonance, while the
$K^{\ast}_{0}(727-263i)$ is not even mentioned in Ref.~\cite{PP},
although evidence for the existence of structure in that
energy region has been reported \cite{BRMDRR86,kappa,Jamin}.
Moreover, a pole in the $S$-matrix is not necessarily
observable as a clear resonance in meson-meson scattering,
as we have argued before.
Furthermore, the $s\bar{s}$ assignment of the $f_{0}(980)$
\cite{f0980,KBRS01,ANS01} hints at the existence of a corresponding,
most probably lower-lying, $n\bar{n}$ structure in $\pi\pi$
scattering \cite{Kunihiro}.

The $f_{0}(1370)$ and $f_{0}(1500)$ resonances have been studied in
many works \cite{KBRS01,ANS01,f01370}, with a diversity of explanations
as to their nature, out of which the above nonet classification is the most
comprehensive.

Of all resonances in Table~\ref{nonets}, the $K^{\ast}_{0}(1950)$
does not seem to be well in place: the general level splittings of some
300 -- 400~MeV do not agree with the jump of 520~MeV
from the ground state to the first radial excitation of the
scalar isodoublet. However, in the analysis of Ref.~\cite{Jamin}
one finds in Table~2 a set of possible singularities (in the
third Riemann sheet) related to the $K^{\ast}_{0}(1950)$
resonance, which all have real parts in the energy region
1.7 -- 1.77~GeV. Moreover, in Ref. \cite{AS97} a central resonance
position of 1.82$\pm$0.04~GeV is reported for this resonance \cite{PP}. 

Furthermore, the $f_{0}(1710)$ is placed at 1.77~GeV in Ref.~\cite{Bugg99},
which, nevertheless, does not alter the above classification,
whereas both the $f_{0}(2020)$ and the $f_{0}(2060)$ need confirmation
and might very well represent the same resonance.
However, if there really exist two resonances in this energy region,
then our classification indicates that one of them must be of 
a nature other than $q\bar{q}$.

In conclusion, one should note that several analyses find {\em too many}
\/$f_{0}$ resonances, whereas in our analysis we {\em lack} \/an $f_{0}(1840)$.
\vspace{1cm}

{\it Acknowledgements.} 
One of us (EvB) wishes to thank the organisers of this conference for the kind
invitation and warm hospitality at Protvino.
We wish to thank Joseph Schechter, H. Noya, David Bugg, Carla G\"{o}bel,
Brian Meadows, Michael Pennington, Teiji Kunihiro, Shin Ishida,
Muneyuki Ishida, Bernard Metsch, and Tadayuki Teshima
for valuable exchanges of ideas.
This work was partly supported by the
{\it Funda\c{c}\~{a}o para a Ci\^{e}ncia e a Tecnologia} under contract
numbers POCTI/\-35304/\-FIS/\-2000 and CERN/\-P/\-FIS/\-40119/\-2000.


\begin{thebibliography}{45}
\bibitem{lQCD}
F.~Butler, H.~Chen, J.~Sexton, A.~Vaccarino, and D.~Weingarten,
Nucl.\ Phys.\ {\bf B430}, 179 (1994)
[hep-lat/9405003];

D.~Weingarten,
Nucl.\ Phys.\ {\bf B215}, 1 (1983);

H.~Hamber, E.~Marinari, G.~Parisi, and C.~Rebbi,
Phys.\ Lett.\ {\bf B108}, 314 (1982);

H.~Hamber and G.~Parisi,
Phys.\ Rev.\ Lett.\  {\bf 47}, 1792 (1981).

\bibitem{qlQCD}  
A.~A.~Khan {\it et al.}  \/[CP-PACS Collaboration],
hep-lat/0105015;

S.~Aoki {\it et al.}  \/[CP-PACS Collaboration],
Phys.\ Rev.\ Lett.\ {\bf 84}, 238 (2000)
[hep-lat/9904012];

Yoshinobu~Kuramashi for the CP-PACS Collaboration,
Proceedings of the Meeting of Particles and Fields of the American
Physical Society (DPF99), Los Angeles, CA, 5--9 Jan.\ 1999
[hep-lat/9904003].

\bibitem{Ricken00}
R.~Ricken, M.~Koll, D.~Merten, B.~C.~Metsch, and H.~R.~Petry,
Eur.\ Phys.\ J.\ {\bf A9}, 221 (2000)
[hep-ph/0008221].

\bibitem{unitarisation}
G.~Fogli and G.~Preparata,
Nuovo Cim.\ {\bf A48}, 235 (1978);

N.~A.~T\"{o}rnqvist,
Annals Phys.\  {\bf 123}, 1 (1979);

Ibid.,
``Bags, Unitarity And Meson Spectroscopy,''
in {\it C81-03-15.12}, HU-TFT-81-20,
Talk given at the l6th Rencontre de Moriond, Les Arcs,
France, March 15--27, 1981.

\bibitem{ccbb80}
E.~van Beveren, C.~Dullemond, and G.~Rupp,
Phys.\ Rev.\ {\bf D21}, 772 (1980)
(Erratum-ibid.\ {\bf D22}, 787 (1980)).

\bibitem{BRMDRR86}                 
E.~van~Beveren, T.~A.~Rijken, K.~Metzger, C.~Dullemond, G.~Rupp, and
J.~E.~Ribeiro,
Z.\ Phys.\ {\bf C30}, 615 (1986).

\bibitem{BRRD83}                 
E.~van~Beveren, G.~Rupp, T.~A.~Rijken, and C.~Dullemond,
Phys.\ Rev.\ {\bf D27}, 1527 (1983);

\bibitem{NUMM}
C.~Dullemond, G.~Rupp, T.A.~Rijken, and E.~van~Beveren,
Comp.\ Phys.\ Comm.\ {\bf 27}, 377 (1982).

\bibitem{BRBW01}
E.~van Beveren and G.~Rupp,
hep-ex/0106077.

\bibitem{Bev84}                 
E.~van~Beveren,
Z.\ Phys.\ {\bf C21}, 291 (1984).

\bibitem{BR99a}
E.~van Beveren and G.~Rupp,
Eur.\ Phys.\ J.\ {\bf C10}, 469 (1999)
[hep-ph/9806246].

\bibitem{Oset}
E.~Oset, J.~A.~Oller, and U.~Mei{\ss}ner,
nucl-th/0109050;

O.~Krehl, R.~Rapp and J.~Speth,
Phys.\ Lett.\ {\bf B390},23 (1997)
[nucl-th/9609013].

\bibitem{DM2}
J.~E.~Augustin {\it et al.}  \/[DM2 Collaboration],
Nucl.\ Phys.\ {\bf B320}, 1 (1989).

\bibitem{ANS01}
V.~V.~Anisovich, V.~A.~Nikonov, and A.~V.~Sarantsev,
hep-ph/0108188.

\bibitem{extended}
E.~van~Beveren and G.~Rupp,
Phys.\ Lett.\ {\bf B454}, 165 (1999)
[hep-ph/9902301];

Ibid.,
Eur.\ Phys.\ J.\ {\bf C11}, 717 (1999)
[hep-ph/9806248].

\bibitem{AS97}
A.~V.~Anisovich and A.~V.~Sarantsev,
Phys.\ Lett.\ {\bf B413}, 137 (1997)
[hep-ph/9705401].

\bibitem{PP}
D.~E.~Groom {\it et al.}  \/[Particle Data Group Collaboration],
Eur.\ Phys.\ J.\ {\bf C15}, 1 (2000).
 
\bibitem{kappa}
Carla~G\"{o}bel, on behalf of the E791 Collaboration,
Proceedings of Heavy Quarks at Fixed Target (HQ2K), Rio de Janeiro,
October 2000, 373-384,
hep-ex/0012009;

Deirdre~Black, Amir~H.~Fariborz, Sherif~Moussa, Salah~Nasri, and
Joseph~Schechter,
Phys.\ Rev.\ {\bf D64}, 014031 (2001)
[hep-ph/0012278];

Deirdre Black, Amir H. Fariborz, and Joseph Schechter,
Proceedings of the YITP Workshop on Possible Existence of the Sigma Meson and
its Implications to Hadron Physics, {\it Sigma Meson 2000},
Kyoto, Japan, 12--14 June 2000 [hep-ph/0008246];

Ibid.,
Proceedings of the International Workshop on Hadron Physics:
{\it Effective Theories of Low Energy QCD}, Coimbra, Portugal, 10--15 Sept.\
1999, AIP Conference Proceedings {\bf 508}, 290 (2000)
[hep-ph/9911387];

Ibid.,
Phys.\ Rev.\ {\bf D61}, 074001 (2000)
[hep-ph/9907516];

M.D.~Scadron,
Proceedings of the YITP Workshop on Possible Existence of the Sigma Meson and
its Implications to Hadron Physics, {\it Sigma Meson 2000},
Kyoto, Japan, 12--14 June 2000 [hep-ph/0007184];

Shin Ishida, Muneyuki Ishida, and Tomohito Maeda,
Prog.\ Theor.\ Phys.\ {\bf 104}, 785 (2000) [hep-ph/0005190]; 

L.~Babukhadia, Ya.~A.~Berdnikov, A.~N.~Ivanov, and M.~D.~Scadron,
Phys.\ Rev.\ {\bf D62}, 037901 (2000)
[hep-ph/9911284];

V.~E.~Markushin and M.~P.~Locher,
Proceedings of the Workshop on Hadron Spectroscopy (WHS 99), Rome, Italy,
8--12 March 1999, {\it Frascati 1999, Hadron Spectroscopy}, 229 (1999)
[hep-ph/9906249];

J.~L.~Lucio Martinez and Mendivil Napsuciale,
Phys.\ Lett.\ {\bf B454}, 365 (1999)
[hep-ph/9903234];

Muneyuki Ishida,
Prog.\ Theor.\ Phys.\ {\bf 101}, 661 (1999) [hep-ph/9902260]; 

Amir H.~Fariborz and Joseph Schechter,
Phys.\ Rev.\ {\bf D60}, 034002 (1999)
[hep-ph/9902238];

Deirdre Black, Amir H.~Fariborz, Francesco Sannino, and Joseph Schechter,
Phys.\ Rev.\ {\bf D59}, 074026 (1999)
[hep-ph/9808415];

Ibid.,
Phys.\ Rev.\ {\bf D58}, 054012 (1998)
[hep-ph/9804273];

J.~A.~Oller and E.~Oset,
Phys.\ Rev.\ {\bf D60}, 074023 (1999)
[hep-ph/9809337];

J.~A.~Oller, E.~Oset, and J.~R.~Pel\'{a}ez,
Phys.\ Rev.\ {\bf D59}, 074001 (1999)
(Erratum-ibid.\ {\bf D60}, 099906 (1999))
[hep-ph/9804209];

Ibid.,
Phys.\ Rev.\ Lett.\  {\bf 80}, 3452 (1998)
[hep-ph/9803242];

Shin Ishida, Muneyuki Ishida, Taku Ishida, Kunio Takamatsu, and Tsuneaki Tsuru,
Prog.\ Theor.\ Phys.\ {\bf 98}, 621 (1997) [hep-ph/9705437]; 

M.~D.~Scadron,
Phys.\ Rev.\ {\bf D26}, 239 (1982).

\bibitem{Jamin}
Matthias~Jamin, Jos\'{e}~Antonio~Oller, and Antonio~Pich,
Nucl.\ Phys.\ {\bf B587}, 331 (2000)
[hep-ph/0006045].

\bibitem{f0980}
F.~De Fazio and M.~R.~Pennington,
hep-ph/0104289;

E.~van Beveren, G.~Rupp and M.~D.~Scadron,
Phys.\ Lett.\ {\bf B495}, 300 (2000)
(Erratum-ibid.\ {\bf B509}, 365 (2000))
[hep-ph/0009265];

E.~M.~Aitala {\it et al.}  \/[E791 Collaboration],
Phys.\ Rev.\ Lett.\  {\bf 86}, 765 (2001)
[hep-ex/0007027].

\bibitem{KBRS01}
F.~Kleefeld, E.~van Beveren, G.~Rupp, and M.~D.~Scadron,
hep-ph/0109158.

\bibitem{Kunihiro}
T.~Kunihiro,
Prog.\ Theor.\ Phys.\ Suppl.\  {\bf 120}, 75 (1995)
[arXiv:hep-ph/9502305].

\bibitem{f01370}
V.~V.~Anisovich, V.~A.~Nikonov, and A.~V.~Sarantsev,
hep-ph/0102338;

R.~Barate {\it et al.}  \/[ALEPH Collaboration],
Phys.\ Lett.\ {\bf B472}, 189 (2000)
[hep-ex/9911022];

D.~Li, H.~Yu, and Q.~Shen,
Mod.\ Phys.\ Lett.\ {\bf A15}, 1781 (2000);

Ibid.,
Eur.\ Phys.\ J.\ {\bf C19}, 529 (2001)
[hep-ph/0011129];

Y.~S.~Surovtsev, D.~Krupa, and M.~Nagy,
Acta Phys.\ Polon.\ {\bf B31}, 2697 (2000)
[hep-ph/0009039];

F.~E.~Close and A.~Kirk,
Phys.\ Lett.\ {\bf B483}, 345 (2000)
[hep-ph/0004241];

D.~Barberis {\it et al.}  \/[WA102 Collaboration],
Phys.\ Lett.\ {\bf B474}, 423 (2000)
[hep-ex/0001017];

T.~Teshima, I.~Kitamura and N.~Morisita,
Nuovo Cim.\ {\bf A103}, 175 (1990);

T.~Teshima, I.~Kitamura and N.~Morisita,
hep-ph/0105107.

\bibitem{Bugg99}
A.~V.~Anisovich {\it et al.},
Phys.\ Lett.\ {\bf B449}, 154 (1999).
\end{thebibliography}
\end{document}